# Getting Around in A College Town: A Case Study of Transportation Barriers faced by International Students at UA

**ABSTRACT**

University students' day-to-day lives largely depend on transportation. Public transit alternatives are not widely available in small-town, rural, and suburban collegiate environments in the United States (U.S.). In this study, an online survey was disseminated among international students studying at The University of Alabama (UA) campus, a predominantly white institution located in Tuscaloosa, Alabama. The objective of this research was to investigate and analyze international student travel experiences in a city with a significant college-affiliated population, as well as to highlight particular transportation issues in the area. The survey results show that international students find it difficult to travel within Tuscaloosa without a personal vehicle, as other modes of transportation including walking, biking, and using public transportation are not always convenient or reliable. The study findings may be of interest to transportation experts, city planners, university administrators, and college students who want to better understand travel-related challenges experienced by international students.

# INTRODUCTION

Over the last decades, there has been a steady growth in the number of international students worldwide and in the United States (Adnett, 2010; Institute of International Education, 2021; Israel and Batalova 2021; Kemp, 2016; West, 2018). Between 1990 and 2014, the number of international students worldwide has quadrupled and reached five million and is expected to increase to eight million in 2025 (West 2018). In the United States (U.S.), the number of international students has grown from 26,000 in 1949-1950 to nearly 1.1 million in 2019-2021 (Institute of International Education, 2021). At the national level, international students are important economically, strategically, and diplomatically, as the movement of international students and scholars around the world helps foster global engagement, ensures the diversity of student bodies on university campuses, promotes cross-cultural understanding, and helps grow enrollment and revenues for higher education institutions (Adnett, 2010 Ruby, 2009; Ward, 2017).

In the United States, the number of international students fell by 1.8% between 2019 and 2020, likely as a result of a series of tightened immigration policies implemented by President Trump's administration as well as safety concerns and travel restrictions related to the global coronavirus disease 2019 (COVID-19) pandemic (Allen & Ye, 2021; Institute of International Education, 2021). However, the U.S. remains the leading host destination in the world for international students with 1,075,496 students enrolled in U.S. higher education institutions in 2020 (Institute of International Education, 2020; Liu & Wang, 2008). Given the benefits that international students bring to the economy and higher education institutions, nearly half of all U.S. universities and colleges currently have strategic international student recruitment plans with specific enrollment targets (Ward, 2017).

# BACKGROUND

## Adjustment Barriers

Since 2015, the number of international students present in the U.S. has consistently topped one million, with the exception of the academic year 2020/2021 when the decline in the international student enrollment was primarily caused by travel restrictions due to the COVID-19 pandemic (Institute of International Education, 2021; Moody, 2021). In the U.S., international students represent 5.5% of the total U.S. higher education population (Institute of International Education, 2019). Existing literature on college student experiences shows that in comparison to their domestic counterparts, incoming international students may experience greater barriers to adjustment and more distress during their initial transition to university life and may report greater academic and career needs (Duong, 2020; Icel & Davis, 2017; Poyrazli & Grahame, 2007). Particularly, some of the more common problems may include lower levels of English language proficiency of incoming international students which may affect their academic performance and social interactions, lack of familiarity with local culture, social norms and expectations, experiences of homesickness, difficulties in finding internships and jobs due to international students' temporary, and non-immigrant student visa status (Duong, 2020; Poyrazli & Grahame, 2007). Past research on the topic suggests that the quality of international student experiences on U.S. college campuses may be closely related to their social participation and cultural assimilation into American society (Baba & Hosoda, 2014; Cemalcilar & Falbo, 2008; Sumer, 2009; Ward 2017).

## Commuting Scenario

In the U.S., the majority of daily trips are made using personal vehicles as a transportation mode: the latest National Household Travel Survey completed in 2017 showed that 87% of daily trips take place in personal vehicles, and 91% of people commuting to work use personal vehicles (Bureau of Transportation Statistics, 2017). Given the U.S.' high dependence on automobiles, especially in small-town, suburban, and rural areas, international students who do not own personal vehicles may face difficulties with transportation and access to essential services and social activities due to the lack of public transportation options. This may significantly impede international students' adjustment to American culture by limiting their access to social interactions and making essential daily tasks, such as grocery shopping, daily commutes to school, participation in after-school activities and social events, more time-consuming and challenging (Poyrazli & Grahame, 2007). Especially in small-towns and suburban and rural areas, where college campuses tend to be spread out while public transportation options tend to be limited, the lack of access to a personal vehicle may result in an increased commute time to school. Previous studies on the topic have shown that longer commute times are negatively correlated with academic performance and students' involvement in extracurricular activities (Alfano & Eduljee, 2013 and Kobus et al., 2015).

University students represent a relatively large portion of the U.S. population. In 2018, college enrollment represented 6.2% of the U.S. population and 41% of the 18- to 24-year-old age group (Bustamante, 2019; National Center for Education Statistics, 2020). Although past research has examined college student activity-travel patterns and travel behaviors along with commuting mode choices in and around university premises (Balsas, 2003; Eom et al., 2009; Sisson & Tudor-Locke 2008; Shannon et al., 2006), there has been a lack of research focusing on transportation-related experiences of those college students that do not have access to personal vehicles. The authors of this paper have not been able to identify any previous studies that would specifically focus on the transportation-related experiences of international students in U.S. colleges.

## Auto-dependency

Delmelle and Delmelle (2012) discussed the reliance on cars by university students and found that car usage in small university towns increases significantly when the walking time from campus to student residence is between 10 and 15 minutes, and then further increases at a higher rate if the walking time exceeds 15 minutes. A survey of commuter students administered on campus at Iowa State University found that 25% of commuter students who live within a two-mile radius from campus prefer to use a personal vehicle to commute to and from school (Zhou et al., 2018). This may be explained by the fact that in a small college town like Ames, where the main campus of Iowa State University is located, the area surrounding the campus might not have all the essential services to meet students' basic needs (such as grocery shopping, healthcare facilities, post office, banking, and financial services and others); thus students may need to travel long distances to access such services. In small-town and suburban areas with limited public transportation options, access to a personal vehicle naturally implies a greater convenience in getting around and a shorter commute time. Likewise, a study conducted at the University of Nebraska Omaha found that out of 234 surveyed commuter students, 77.7% traveled to campus alone on a daily basis using a personal vehicle (Grant, 2008). The negative consequences of such a trend include high volumes of automobile traffic and high demand for parking (Daisy et. al., 2018).

Few research studies have focused on the issue of auto-dependency among university populations (Barla et al., 2012; Shannon et al., 2006). When it comes to the issue of auto-dependency among international students, hardly any research on the topic has been conducted.

The objective of this study was to contribute to academic literature about international students' transportation-related experiences and document issues of auto-dependency that international students may face on college campuses located in small-town, suburban, and rural settings.

## METHODOLOGY

### Study Context

This study was conducted at The University of Alabama's (UA) flagship campus in Tuscaloosa, Alabama, and was based on the results of an online survey administered among UA international students. The objective of the study was to examine international students' experiences when it comes to getting around a city that hosts a significant college-affiliated population and to highlight issues related to transportation options in the area. Based on the study objective, following research question was investigated: *What transportation-related barriers do international students at UA face?*

 UA was recently ranked 159th out of 1,300 colleges and universities in the U.S. for popularity among international students (OIRA, 2019). According to the statistical data provided by the *Students by Level and Geographic Origin Report* from the Office of Institutional Research and Assessment (OIRA), in the fall of 2018, there were 1,219 international students from 76 countries enrolled in UA. These students represented 3% of the total 38,392 students enrolled at UA during that term (OIRA, 2018).

Tuscaloosa is a college town with a population of 101,129 people (U.S. Census Bureau, 2019). While the UA campus itself is a pedestrian- and bicycle-friendly environment and offers a variety of free public transportation services to its community during the academic year, the City of Tuscaloosa currently has very few public transportation options. The business district and downtown areas of the city are walkable and bicycle-friendly. However, other parts of Tuscaloosa may be challenging to access without a personal vehicle. Outside of Tuscaloosa's downtown area, there is a general lack of pedestrian and bicycle facilities (such as sidewalks and bike lanes) and most of the existing walking and biking infrastructure is not ADA-compliant (The City of Tuscaloosa, 2019). The Tuscaloosa County Parking and Transit Authority (TCPTA) provides transportation services via bus, van, and trolley that serve only a limited number of areas within the city limits, between 5 am - 6 pm on weekdays, with no weekend service available (Tuscaloosa Transit Authority, n.d.). As a result, international students at UA who do not have access to a personal vehicle may find it challenging to get around the city. That said, UA runs a free shuttle service to several department stores and retail locations near campus for limited Saturday hours as well as an on-call service on Sundays (also for limited hours). Furthermore, UA provides a rideshare shuttle service (348-Ride) on weeknights to serve a limited-service area with daily operations from 9 pm-12 am. It is a convenient way for students to safely return to their places of residence later in the night. The findings of this study may be generalized to college towns of similar size across the U.S. As such, the outcomes of this study may be of interest to transportation researchers, city planners, university administrators, and students motivated to learn more about the travel-related challenges experienced by international students in the U.S.

**Participants**

The study's target population was the population of international students registered at UA in the Fall 2018 semester. The participation was voluntary, and respondents were not compensated for participating. Based on a 95% confidence level and 10% error margin, the ideal sample size for the research would be 90. We recruited 110 participants, which represented approximately 9% of the target population. Survey participants represented 35 countries around the globe. Figures 1 and 2 illustrate the survey respondents' distribution by enrollment year and age range, respectively.

Figure 1

*Survey Respondent Year of Enrollment*

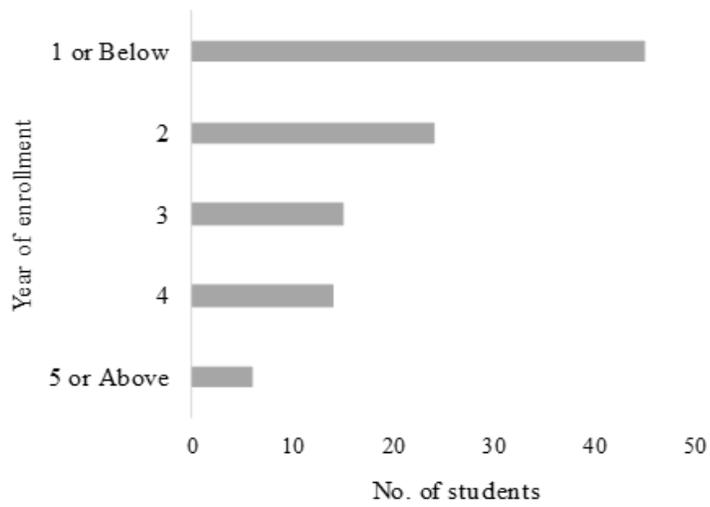

Figure 2

*Survey Respondent Age Group at UA*

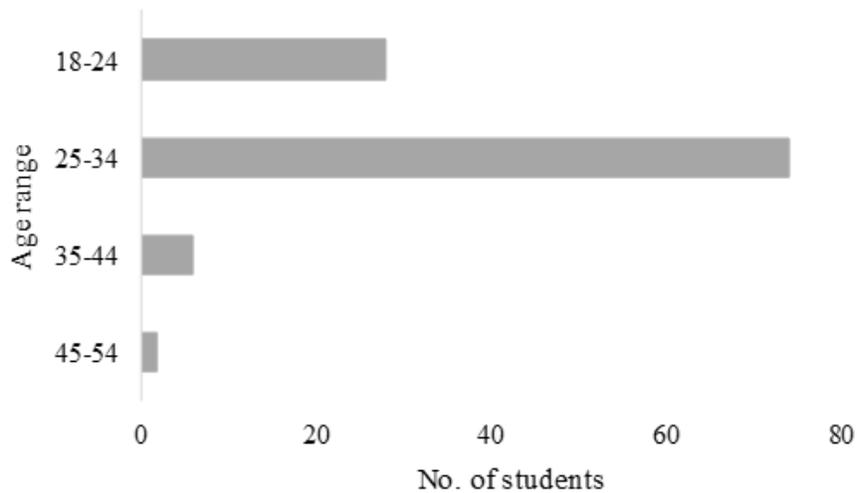

The following table (Table 1) provides a comparison of the study sample with the international student population enrolled at UA for the Fall 2018 semester (OIRA, 2018). Women were disproportionately represented in the survey; specifically, there were 47.27% of participants who identified as women. At the same time, the proportion of the UA international student population that self-identified as women averaged 29.04% in the Fall 2018 semester. It is important to note that the University of Alabama's statistical data on international students only includes two gender categories ("male" and "female") while the data is based on students' self-reported responses. Therefore, the data may not reflect all other gender categories that international students enrolled at the University of Alabama may hold (such as transgender, non-binary, and others).

Most of the participants (75.40%) were graduate students in doctoral or master's programs, and less than 20% were undergraduate students. The remaining students were in non-degree seeking programs such as UA English Language Institute (ELI) programs. The percentage of graduate student participants also differed drastically from that of the international graduate student population. Furthermore, the sample included a larger proportion of students who identified Asia as their region of origin than the overall international student population (63.30% vs. 49.06%). In contrast, students that identified the Middle East and North Africa as their region of origin were underrepresented in the survey compared to the overall international student population at UA (11.10% vs. 27.81%).

Table 1.

*Comparison between International Student Population and Study Sample*

| | Total Population of International Students Enrolled at UA as of Fall 2018 | Survey Respondents |
|---|---|---|
| **Gender** | | |
| Male | 70.96% | 51.82% |
| Female | 29.04% | 47.27% |
| Transgender/Gender Nonconforming | - | - |
| Prefer not to respond | - | 0.91% |
| **Degree Level/student status** | | |
| First-year | 13.20% | 5.50% |
| Sophomore | 6.73% | 3.60% |
| Junior | 12.96% | 4.50% |
| Senior | 15.18% | 3.60% |
| Master's | 8.37% | 10.90% |
| Doctoral | 35.68% | 64.50% |
| Others | 7.88% | 7.30% |
| **Region of Origin** | | |
| Asia | 49.06% | 63.30% |
| Europe | 10.01% | 8.26% |
| Latin America | 6.81% | 9.17% |
| Middle East & North Africa | 27.81% | 11.01% |
| North America | 2.05% | 0.92% |
| Oceania | 1.31% | 0.92% |
| Sub-Saharan Africa | 2.95% | 6.42% |

**Procedures**

Student survey is a popular method to capture the feedback of the student population (Qualtrics, n.d.-a). The perceptions of college students have been studied in the past using both paper surveys (Skeeter et al., 2019) and internet-based questionnaires (Grant, 2008; Barla et al., 2012; Shannon et al., 2006). Online survey questionnaires are cost-effective for many institutions and handy for tech-savvy participants like college students (Carini et al., 2003; Shannon et al., 2006). Conducting a student perception survey include planning, creating, administering, discussing survey results, and acting on survey findings using a survey questionnaire containing close-ended questions, likert scale, multiple choice, and open-ended questions (Qualtrics, n.d.-a). A survey questionnaire was designed in Qualtrics following these criteria. Before it was distributed, the questionnaire was field reviewed and evaluated by research associates at a prominent research institute. Based on the feedback received, the authors revised the survey several times before it was finalized. Since the research included human subjects, authorization from the University of Alabama's Institute Review Board was obtained before disseminating the survey. The International Student & Scholar Services (ISSS) office at UA agreed to assist with distributing the survey to international students who were enrolled at UA at the time of the survey. ISSS distributed the survey via a link posted to their monthly emails and their official social network accounts including Facebook, Twitter, and LinkedIn. The survey link included a QR code that allowed participants to scan the code through their mobile phones. The survey was published online on March 4, 2019 and was closed on April 11, 2019.

**Measures**

The survey questionnaire consisted of 36 questions (see Appendix A). Among these questions, 23 were main questions and 13 were branching questions (branching questions send respondents down different paths in the survey, depending on how they answered screening questions). The majority of survey questions were multiple choice; however, there were a few open-ended, Likert scale, and dropdown questions. The survey questions were divided into five categories: demographic, student background, ownership of personal vehicle, travel information, and opinions. Table 1 shows completion rates for each question along with the question type. Even though it may be confusing for readers when the sample size varies from question to question, including partial replies reduces the bias of subject salience and enhances the sample size for initial questions (Henning, 2021). The number of responses for each question was used to compute descriptive statistics for that specific question following the approach of Skeeter et al. (2019). For instance, the stated percentage of age was estimated as a percentage of 110, whereas the present student level on campus was calculated as a percentage of 109 (Table 2).

Table 2

*List of Survey Questions and Corresponding Number of Responses and Question Type*

| Questions | Number of Responses (n) | Completion Rate | Question Type |
|---|---|---|---|
| **Demographics** | | | |
| Gender | 110 | 100.00 | Multiple Choice |
| Age | 110 | 100.00 | Multiple Choice |
| Country of Residency | 103 | 93.64 | Dropdown |
| Current Student Level on Campus | 109 | 99.09 | Multiple Choice |
| Is Your Current Address within the Grid Map? | 107 | 97.27 | Multiple Choice |
| Grid Number Based on Current Address* | 92 | 83.64 | Dropdown |
| Current Address Outside of Grid Map* | 14 | 12.73 | Multiple Choice |
| On-Campus Job Employment Status | 106 | 96.36 | Multiple Choice |
| On-Campus Job Title* | 76 | 69.09 | Multiple Choice |

| Questions | Number of Responses (n) | Completion Rate | Question Type |
|---|---|---|---|
| **Student Background** | | | |
| Year of Enrollment at UA | 103 | 93.64 | Open-ended |
| Driver's License Status in Home Country | 105 | 95.45 | Multiple Choice |
| Popular Travel Mode in Home Country | 102 | 92.73 | Multiple Choice |
| Preferred Travel Mode in Home Country | 102 | 92.73 | Multiple Choice |
| **Vehicle Ownership** | | | |
| Do You Own A Personal Vehicle? | 105 | 95.45 | Multiple Choice |
| Status of U.S Driver's License | 110 | 100.00 | Multiple Choice |
| Did You Purchase A Personal Vehicle Before or After Arriving in Tuscaloosa?* | 63 | 57.27 | Multiple Choice |
| When Did You Purchase A Personal Vehicle After Arriving in Tuscaloosa?* | 46 | 41.82 | Multiple Choice |
| Payment Option for Personal Vehicle Purchase* | 60 | 54.55 | Multiple Choice |
| Using On-Campus Parking Permit* | 62 | 56.36 | Multiple Choice |
| **Travel Information** | | | |
| Preferred Travel Mode in Tuscaloosa | 102 | 92.73 | Multiple Choice |
| Places Most Visited Soon After Arriving in Tuscaloosa | 101 | 91.82 | Multiple Choice |
| Travel Mode for Getting to Places Soon After Arriving in Tuscaloosa* | 102 | 92.73 | Multiple Choice |
| Do You Stay on Campus Past 9 pm? | 102 | 92.73 | Likert-Scale |
| Preferred Travel Mode for Getting Home After 9 pm* | 87 | 79.09 | Multiple Choice |
| Have You Ever Used Public Transportation System in Tuscaloosa (Besides the UA Bus Service)? | 102 | 92.73 | Multiple Choice |
| How Frequently Do You Use Public Transportation System in Tuscaloosa (Besides the UA Bus Service)?* | 25 | 22.73 | Likert-scale |
| Have You Ever Used Services like Uber, Lyft, or Taxi in Tuscaloosa? | 102 | 92.73 | Multiple Choice |
| **Opinions** | | | |

| Questions | Number of Responses (n) | Completion Rate | Question Type |
|---|---|---|---|
| Do You Prefer Using Public Transit Over Personal Vehicle? | 102 | 92.73 | Likert-Scale |
| Perceptions of Existing Public Transit Services in Tuscaloosa | 102 | 92.73 | Multiple Choice |
| Perceptions of Using Uber, Lyft, or Taxi Services in Tuscaloosa* | 53 | 48.18 | Multiple Choice |
| Perceptions of Conveniences of Using Uber, Lyft, or Taxi in Tuscaloosa* | 27 | 24.55 | Multiple Choice |
| Perceptions of Inconveniences of Using Uber, Lyft, or Taxi in Tuscaloosa* | 6 | 5.45 | Multiple Choice |
| Perceptions of Existing Public Transportation System in Tuscaloosa | 102 | 92.73 | Multiple Choice |
| Do You Think That It Is Challenging to Get Around Tuscaloosa Without A Personal Vehicle? | 102 | 92.73 | Likert-Scale |
| Additional Comments | | | Open-ended |

**Branching Questions

This study collected quantitative data from closed- and open-ended survey questions presented to international students at the University of Alabama. Descriptive statistics are the fundamental measurements that are used in the process of describing survey data since they consist of summative descriptions of individual variables and the accompanying survey sample (CVent, 2019; Qualtrics, n.d.-a; Torchim, n.d.). Whereas, the traditional approach for analyzing responses to open-ended questions is quantitative content analysis (Züll, 2016). Descriptive statistics with quantitative content analysis include the opportunity to learn more about participants' experiences and enrich the findings of the study. Hence, the study employed descriptive statistics to analyze the close ended questions and quantitative content analysis to analyze the open-ended questions. The survey responses were exported to Microsoft Excel from Qualtrics. Each author carefully examined the open-ended question answers to determine how often specific categories of information came up. The authors then went through these categories and summarized the transportation-related barriers that international students at the UA encounter.

## RESULTS

Researchers used quantitative content analysis to summarize and explain the open-ended questions, and descriptive statistics to evaluate the remainder of the survey data. The results obtained from the survey highlight international students' shared experiences related to getting around in a mid-sized college town.

### High Automobile Ownership Rates

The survey allowed gathering information about the number of respondents who held a U.S. driver's license, the number of respondents who owned a personal vehicle in Tuscaloosa, and how soon those survey respondents that owned a vehicle, purchased it after they arrived in the U.S. The results were summarized in Table 3.

More generally, the following comments summarize the extent to which respondents perceived their need for access to a personal vehicle in Tuscaloosa:

- Respondent 1, a master's student, said that "I usually walk to close destinations on daily basis, but I feel so strange since I am the only one walking on pedestrians. I have never been to this kind of city where everyone is using personal vehicle, and nobody is walking."

- Respondent 2, a senior student mentioned that "Car is necessary. Can't go anywhere without personal vehicle."

- Respondent 3, a Ph. D. student said that, "I experience a one year without car and it took time and money to get around town."

Table 3

*Summary of Survey Respondent Vehicle Ownership Situation in Tuscaloosa*

| Do you have a U.S. driver's license? | | |
|---|---|---|
| | No. of responses | Percentage |
| Yes | 63 | 60% |
| No | 25 | 23.80% |
| Learner's Permit | 12 | 11.40% |
| International driver's license | 5 | 4.80% |
| **Do you have a personal vehicle?** | | |
| | No. of responses | Percentage |
| Yes | 63 | 60% |
| No | 42 | 40% |
| **How long after arriving in Tuscaloosa, did you purchase a personal vehicle?** | | |
| | No. of responses | Percentage |
| Within a week | 1 | 2.17% |
| Within a month | 6 | 13.04% |
| Within 3 months | 6 | 13.04% |
| Within 6 months | 8 | 17.39% |
| Within 1 year | 8 | 17.39% |
| Within 1.5 years | 12 | 26.09% |
| 2 years and above | 5 | 10.87% |

**Lack of Reliable Public Transportation**

Generally, many survey respondents found Tuscaloosa to be highly auto dependent and reported the lack of reliable public transportation options to be a major hindrance in getting around the city. The survey found that public transportation was not popular among international students. Among 102 students, upwards of 75.5% students reported that they never used any public transportation within Tuscaloosa except for the free transportation services offered to the campus community by UA. Within this group, nearly 46% were not aware of the availability of any existing city bus service in Tuscaloosa whereas another 35.3% found the service to be highly infrequent and unreliable. Overall, 67% of the respondents expressed dissatisfaction with the existing transportation system in Tuscaloosa.

We used a five-point Likert scale to ask respondents to rate their preference for using public transportation over personal vehicles; the options ranged from "definitely yes" to "definitely not". Thirty-five percent (35.3%) answered "definitely yes", 12.8% answered "probably yes", 11.8% answered "might or might not", 19.6% answered "probably not", and 20.6% answered, "definitely not". When responding to a different question in the survey, nearly 75.5% out of the 102 respondents said that they had never used Tuscaloosa's public transportation services. The following open-ended responses by survey participants reflect the extent to which the respondents perceived a lack of reliable public transportation in Tuscaloosa:

- Respondent 4, a Ph. D. student mentioned that, "Public transportation doesn't work on weekends. I think on Saturdays it only runs up to noon and it doesn't reach all parts of the city and residential areas. The routes are so limited and depending on where you live, it will take you a long time to get to where you want to go because it has to go all over before it can get to where you need to be. It is very time-consuming."

- Respondent 5, a Ph. D. student said that, "The public transportation only takes people to certain distance; however, a lot of places are located beyond that and are somewhat necessary for student living such as Walmart, Water service etc."

**Limited Access to Facilities**

The survey inquired about the places international students visited most frequently upon their arrival in Tuscaloosa keeping in mind that typically, international students would not have access to a personal vehicle immediately after their arrival (this question offered the opportunity to choose several responses categories). Out of 110 respondents, nearly 40% went to department stores like Walmart, Target, etc. 20.9% went to utility company offices, 16% to the local driver's license office, 11.9% went to the court to obtain their social security documents, 7.8% went to the local post office, and another 4.1% went to other local places. In order to reach these places, 64.7% of respondents noted that they asked their friends to drive them while another 32.4% used a personal vehicle.

The survey provided respondents with a map which helped the researchers gain a better understanding of how far most students lived from the campus area. Particularly, we learned that a total of 75% of the respondents were located just a few miles from the UA campus. Whereas, many department stores, utility company offices, and the local driver's license office among other places are located farther away from the campus area and thus, not within a reasonable walking or biking distance.

Additionally, in the survey, students were asked to share their opinions about the convenience of existing ride-sharing services provided in Tuscaloosa such as Uber and Lyft. Among 32 respondents, 24 said that these services were highly convenient since they were easily available and another 7 suggested that they were safe to ride. More generally, the following comments highlighted the respondents' perception of their limited access to facilities:

- Respondent 6, a Ph. D. student mentioned that, "Without car, somebody should spend whole day to get somewhere and come back home."

- Respondent 7, a Ph. D. student stated that, "Getting anywhere other than school is almost impossible without a car. For example, other than Publix, there isn't any grocery store that's a walking distance from campus."

- Respondent 8, a Ph. D. student said that, "You are kind of confined to the campus bubble. especially if you can't walk or bike far either"

**Perceived Safety Concerns**

The results of the survey highlighted safety concerns among international students regarding getting around in Tuscaloosa without a private vehicle. Female students especially emphasized such concerns as they shared that walking or biking outside the UA campus area made them feel vulnerable and unsafe. The survey found that out of 19 respondents, 42.1% of females preferred walking back home from campus after 9 p.m., and out of 7 respondents, no females preferred to bike back home from campus after 9 p.m. Additionally, out of 14 respondents, females were 28.6% more likely to ask a friend to drop them home after leaving campus past 9 p.m. The following comments by a female graduate student and a senior year female student reflect their views on perceived safety around campus:

- Respondent 9, a Ph. D. student stated that, "It is impossible to live without a car. It is unsafe walking or biking around the city if you don't live on campus."

- Respondent 10, a senior student said that, "Before I bought my car, I could not go to any places because it was not easy and convenient for me. Actually, it is not safe enough to walk in the city at night."

**High Financial Burden**

Purchasing a personal vehicle is a significant expense for any student. With this understanding in mind, the survey asked respondents whether they owned a personal vehicle and how they purchased their vehicle. Out of 60 respondents, 70% reported that they purchased their car with cash, 8.3% reported taking out bank loans, and another 1.67% leased their vehicle.

Furthermore, students reported feeling a significant amount of financial burden related to traveling using ride-sharing company vehicles to get around the city. Among those who had used ride-sharing services, 85.7% believed that ride-sharing services were very costly. More generally, the following comments summarize the extent to which respondents perceived financial constraints to be a barrier within the broader conversation about automobile dependence in Tuscaloosa:

- Respondent 11, a Ph. D. student said that, "Not everyone knows driving or can afford buying a car when they come to the USA for the first time."

- Respondent 12, a Ph. D. student mentioned that "Taxies are costly. Comparatively cheap stores are far from campus. If you don't have car, you have to depend on other which is not convenient. If you don't depend, it won't be convenient for the pocket."

- Respondent 13, "If I'd not have one, I will need to use Uber or taxi and they cost a lot of money."

**DISCUSSION**

**Difficulties in getting around**

This study revealed that most UA international students that participated in the survey found it difficult to travel around Tuscaloosa, Alabama without a personal vehicle. This is consistent with the previous research showing U.S. residents' relatively high level of auto-dependency and transportation-related challenges experienced by college students that do not have access to a personal vehicle (Bureau of Transportation Statistics, 2017; Poyrazli and Grahame, 2007; West, 2021). The responses of our study participants have shown that most international students arrive in Tuscaloosa without a car but decide to purchase a personal vehicle soon after their arrival. Particularly, over half of those UA international students that reported they owned a car when they took the survey, said that they purchased their cars within the first 6 months after they arrived in Tuscaloosa. Yet, it is important to realize that purchasing a personal vehicle may not be feasible for those international students who have limited financial resources. Lower-income international students similar to lower-income American college students may be disproportionally affected by limited access to transportation which, as explained earlier in the paper, may lead to increased commute time (and therefore, reduced time available for academic and extracurricular activities). Yet in the case of international students, the negative impact of limited transportation options maybe even stronger than that for their American counterparts because for international students, the issue of limited transportation access is combined with language and cultural barriers, lack of ties to the community and the absence of a support system such as family and friends (who may occasionally be able to give a ride, let the student borrow a car, or offer another kind of support).

Most survey respondents, who attended classes and worked on the UA campus, owned a personal vehicle but chose not to hold a UA parking permit. This demonstrates that within the UA campus, most survey respondents were comfortable with using transportation options other than a personal vehicle (such as biking, walking, and UA transportation services). However, the same respondents stated that traveling outside the UA campus was not convenient without a personal vehicle. The respondents reported that many areas in Tuscaloosa were not designed for walking or biking due to the lack of sidewalks and bike lanes. Indeed, students who live within walking distance from the UA campus and mostly rely on walking and public transportation as their main transportation modes may find it difficult to access essential services that are located further away from the campus. Additionally, walking and biking in Tuscaloosa were found to be associated with safety concerns, especially among female respondents.

The survey also showed that over three-quarters of the respondents have never used public transportation in Tuscaloosa other than the UA campus transportation services. A little over 40% of the survey respondents stated that they "probably" or "definitely" would not use public transportation in Tuscaloosa. The respondents who had the experience of using public transportation services in Tuscaloosa reported that these services were not always convenient since they did not cover most of Tuscaloosa's neighborhoods. Additionally, the respondents stated that the schedule of public transportation service routes was limited, especially on holidays and weekends. While many international students found public transportation services provided by the City of Tuscaloosa inconvenient and limited, the participants stated that using Uber or Lyft on a regular basis was costly.

**Challenges of vehicle ownership**

For those international students who decide to purchase a personal vehicle in the U.S., doing so may be less affordable than for their American counterparts. Individuals who are not U.S. citizens or permanent residents may often encounter challenges when trying to obtain credit loans from car dealerships and banks. Indeed, in most cases, international students may find themselves in a situation where paying for a car purchase in full in cash may be the only feasible option (Vanderbilt University ISSS, n.d., International English Institute, 2017). This is consistent with the finding of our study: 71.21% of participants reported that after they arrived in the U.S., they paid cash for their cars in full. This trend was nearly opposite to what is generally observed among the U.S. population; between the first quarter of 2017 and the second quarter of 2020, the share of new vehicles purchased in the U.S. using financing as a payment method was between 85.5% and 87.9% (Statista.com, 2020).

Furthermore, it may not always be practical and feasible for an international student to own a personal vehicle if the purpose of the student's stay is to participate in a short-term academic program (such as the UA English Language Institute programs that usually last between one-twelve months and offer reading, writing, oral communication, and grammar courses for international students). At the same time, living in an automobile-dependent city without a personal car may pose significant difficulties and reduce the quality of students' experiences even if the duration of a student's stay is no longer than a few months.

**Socioeconomic mobility concern**

Every year, hundreds of thousands of international students arrive in the U.S. in search of education with the expectation that such academic experience would help them improve their future employment opportunities and have a positive effect on their socioeconomic mobility. The high quality of U.S. higher education and its perceived value on the international labor market improved access to job opportunities in the U.S., and a positive impact on the socio-economic mobility of the international alumni of U.S. higher education programs are some of the factors that make international students choose the U.S. as a destination for their study abroad experience (Israel and Batalova, 2021). Socioeconomic mobility refers to the concept of climbing the socioeconomic ladder from childhood to adulthood and could be defined as the difference in an individual's income, wealth, or occupation in adulthood from that of the individual's family when he/she was a child (Slaughter-Acey et al., 2016).

Past research has shown that individuals' transportation mobility (i.e., ease of travel from one place to another) impacts their socioeconomic mobility and that there is a strong relationship between socioeconomic mobility and transport disadvantage (Hine, 2012). Lower levels of access to transportation tend to limit access to employment opportunities, shopping, services, health facilities, and recreational activities (Jansuwan et al., 2013; Hine, 2012). In the context of a college student's life, limited access to reliable transportation options may imply longer commute times to, and from school, grocery stores, recreational facilities, and other essential services. Additionally, limited access to transportation may make it challenging to access professional development events and job interviews as well as restrict one's ability to hold off-campus jobs and internships for an extended period of time. These combined disadvantages can have a negative impact on an individual's quality of life and also reduce their chances of professional advancement after graduation. Auto-dependency in rural and suburban college campuses is a burden on most students that do not have access to a personal vehicle, and such a disadvantage is even more pronounced in the case of low-mobility individuals including older adults, low-income individuals, and people with disabilities (Jansuwan et al., 2013). Research has shown that among the different groups of low-mobility individuals, individuals with disabilities are more likely to rely on public transportation than on private transportation (Jansuwan et al., 2013). Therefore, international students with disabilities arriving at rural or suburban campuses in the U.S. may find themselves in an especially challenging situation where campus transportation options are limited, and the use of a private vehicle is not feasible.

## FUTURE DIRECTIONS

Although public transportation options in Tuscaloosa are limited compared to large metropolitan areas, survey responses showed that many international students were not familiar with all of the services currently available in the city. This made it harder for students who didn't have a car to get around. Therefore, it may be important for the UA ISSS to make additional efforts to help incoming students learn more about all of the transportation options available in Tuscaloosa. Additionally, it may be helpful to expand the existing public transportation network in Tuscaloosa in order to serve more communities. For example, Iowa State University in collaboration with the City of Ames provides a bus service called CyRide that serves multiple communities outside of the ISU campus. The ISU Student Government's general funds cover 45% of the operation and maintenance cost of CyRide (Zhou & Yin, 2018). Similar public transportation partnerships between universities and municipal governments may be helpful in other college towns across the country.

## LIMITATIONS

It is important to acknowledge a few limitations of the study. Particularly, when considering the findings of the study, one must note the demographic bias of the sample, if compared to the overall population of international students enrolled at UA in the fall 2018 semester and the relatively small size of the sample (110 observations). While invitations to participate in the survey were sent to all foreign students regardless of their demographic status, women and graduate students were more likely to respond.

Although the gender breakdown of international students enrolled at UA during the Fall 2018 semester was 71% men vs. 29% women, the gender breakdown of survey participants was 52% men vs. 47% women. While the previous study indicated that males were more likely than females to complete a survey on the Web (Smith and Leigh, 1997), Underwood et al. (2000) showed that college women were more likely than males to reply to any survey technique, including the Web. However, the authors think that perceived safety concerns may be one of the main reasons why female students were represented in the survey more than expected. Another notable inconsistency in the survey response was the larger number of graduate-level students (75%) participating in comparison to the percentage (44%) of international graduate students enrolled at UA during the Fall 2018 semester. According to Park et al. (2008), undergraduate students are more concerned with the technical aspects of a survey (i.e. survey duration, convenience, etc.), but graduate students are more likely to respond according to content. While investigating the characteristics that influence student survey participation and method of completion, Hunt-White (2006) found that students with a higher degree are more likely to participate in a survey, however, this probability may rise if the survey is conducted online. As a result, the authors perceive that the survey topic may have piqued the interest of graduate students. Additionally, students show more or less psychological reactions based on where they come from (Poyrazli & Grahame, 2007). For example, compared to European students, students from Asia, Central/South America, and Africa report more acculturative stress (Yeh & Inose, 2003). This may explain why Asian, Latin American, and Sub-Saharan African students were overrepresented in the survey.

An ideal representative sample with the characteristics identical to the population from which it was chosen may be unattainable (Education Development Center, 2018). Given the demographic bias of the sample, it is possible that the recorded survey responses may not be representative of the population of all international students at UA. However, the authors think that, if interpreted qualitatively, the findings of this study could meaningfully contribute to the discussion about ways to improve international student experiences in the U.S.

The authors opted to conduct the study with the sample size obtained considering that a significant amount of information would be lost if a larger sample size is consistently insisted upon (IPCT, Robin Hill). Moreover, numerous survey researchers have started to cast doubt on the generally accepted belief that low response rates produce biased findings (Curtin, Presser, & Singer, 2000; Groves, 2006; Keeter, Miler, Kohut, Groves, & Presser, 2000; Massey & Tourangeau, 2013; Peytchev, 2013). Given the gap in current research about college students' experiences (especially, international students' experiences) related to transportation options in college towns, the results of this study may be valuable to researchers, university workers, stakeholders, and policymakers that work to improve international students' experiences at the U.S. college campuses. Additionally, the survey methodology applied in this study may be helpful in the development of a larger-scale study about international students' experiences with transportation options serving many U.S. college campuses located in small-town, rural, and suburban settings.

## CONCLUSION

Transportation and mobility-related challenges faced by international students are complex and often interconnected with a multitude of social and psychological factors. When arriving in the U.S. for the first time to pursue their studies, international students may often face challenges that are typical for those relocating to a new country, including the lack of a social support system such as family and friends in the new country of residence, limited access to a personal vehicle, difficulties in communication due to language and cultural barriers, and others. These challenges, combined with prejudice, negative stereotyping, and discrimination that may sometimes be experienced by international students, may negatively affect international students' overall experiences in the U.S. and make them feel unwelcome and marginalized. Careful and thoughtful transportation planning in U.S. college towns may help alleviate some of the international students' challenges, enhance the opportunities to find friends, build a social support network and succeed academically through improved access to essential services (such as shopping, healthcare services, financial institutions, and others), access to campus outside of business hours for extracurricular activities, social events, tutoring, sports, and recreation, etc. (Ward, 2017). Therefore, improving international students' transportation-related experiences may greatly benefit their physical, psychological, and social well-being.

Furthermore, improving the experiences of international students is strategically important to the U.S. higher education system. International students tend to be high-ranked students in their home countries even though some of them may initially have lower levels of English language proficiency when enrolling in U.S. institutions, compared to their American counterparts (Wu, Garza, and Guzman, 2015). International students enrich the cultural diversity of U.S. college campuses and provide opportunities for American faculty, students, and the communities where they live to enhance their cultural awareness and sensitivity and learn about different cultures and traditions (Wu, Garza, and Guzman, 2015). Lastly, international students represent a significant investment for U.S. universities and campus communities from an economic standpoint because international students may help universities and nearby communities generate higher revenues from international students' academic, living, travel, and entertainment expenses (Wu, Garza, and Guzman, 2015).

Some possible ways to improve international students' transportation-related experiences include providing informational support to incoming international students and their families through pre-arrival outreach initiatives and on-campus orientation for international students. Thorough planning needs to be conducted at the university level to develop strategies for cost-effective and possibly, cost-neutral methods to provide transportation and transition support to international students after their arrival. Additionally, administering additional transportation services at the beginning of an academic semester (such as airport pick-up and shared transportation services) could help international students without personal vehicles perform necessary tasks upon their arrival on campus and facilitate the adjustment process (Ward 2017).

Lastly, more research needs to be done to understand students' experiences with transportation options in college towns. This study identified that student commuters are often underrepresented in travel surveys (Volosin, 2014). Along with international students and scholars, students and employees from low-income communities as well as persons with disabilities may be among campus populations that do not have reliable access to a personal vehicle and therefore, may experience challenges due to limited public transportation options in small-town, rural and suburban settings.

**Appendix A**

**Survey Questionnaire**

1. Your Gender

   - Male

   - Female

   - Transgender/ Gender Nonconforming

   - Prefer not to respond

2. Please select your age range

   - Under 18

   - 18 - 24

   - 25 - 34

   - 35 - 44

   - 45 - 54

   - 54 or older

3. What is your country of residence/home country?

▼ Afghanistan (1) ... Zimbabwe (1357)

4. What is your current status on campus?

- Freshman

- Sophomore

- Junior

- Senior

- Master's Student

- PhD Student

- Other __________________________

5. Please select your age range

- Under 18

- 18 - 24

- 25 - 34

- 35 - 44

- 45 - 54

- 54 or older

6. What is your current status on campus?

- Freshman

- Sophomore

- Junior

- Senior

- Master's Student

- PhD Student

- Other ______________________________

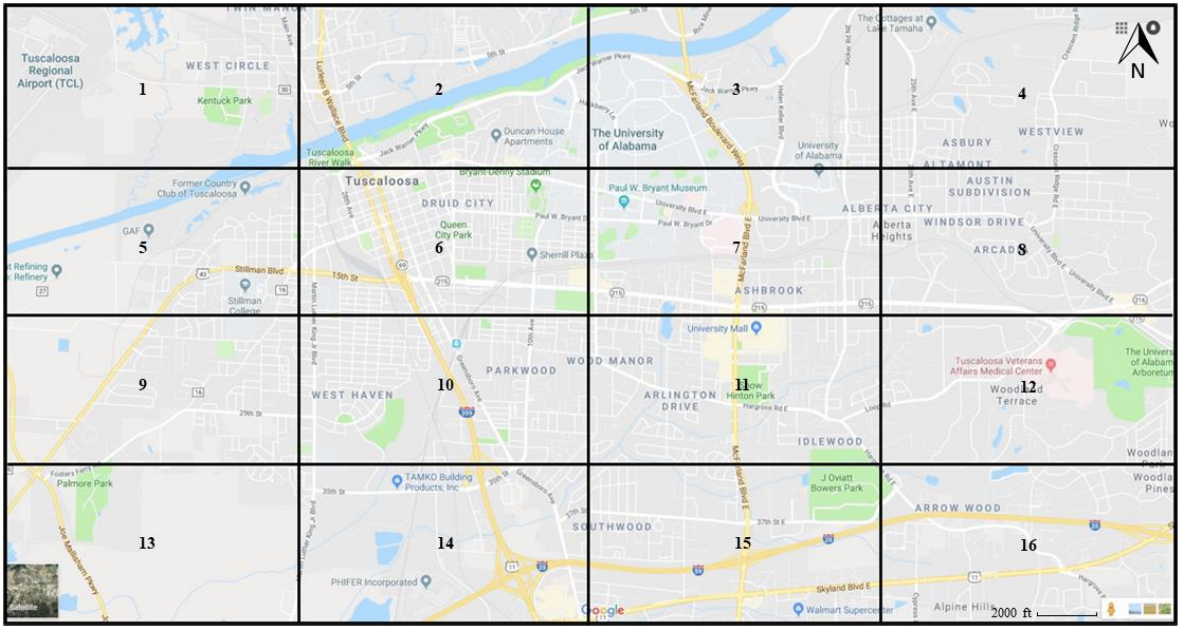

7. Do you live within this grid map?

- Yes

- No

Please select a grid number that includes your present address

- ▼ 1 (1) ... 16 (16)

Please specify (if 2, 3 is selected)

- North side of the river

- South side of the river

Please specify where do you live (If 7 is No) -

- North side of the river

- South of Skyland boulevard

- East side of the grid map

- West side of the grid map

- Any other city of Tuscaloosa County (Please Specify)

  ______________________________________________

8. Do you have an on-campus job?

   - Yes

   - No

   Please Specify

   - Graduate Teaching Assistant

   - Graduate Research Assistant

   - Grader

   - Student Worker (Hourly Position)

   - Tutor

   - Self-employed

   - Other ______________________________

9. What is your year of Enrollment in UA?

   ______________________________________

10. Did you have a driver's license back in your home country?

    - Yes

    - No

11. What is the available and popular transportation system to travel within your city back in your home country?

- Public transit (Bus, Train, Subway etc.)
- Personal vehicle
- Uber, Lyft, Taxi etc.
- Bike
- Motorbike
- Other ____________________________

12. What did you use more frequently back in your home country?

- Public Transit (Bus, Train, Subway etc.)
- Personal Vehicle

13. Do you have US driver's license?

- Yes
- No
- I have learner's permit
- I have international driver's license

14. Do you have a car in the US?

- Yes
- No

15. When did you buy your personal car in the US? (if 14 is Yes)

- Before arriving in Tuscaloosa

- After arriving in Tuscaloosa

16. When did you buy your personal car after your arrival in Tuscaloosa? (If 15 is After arriving in Tuscaloosa)

- Within a week
- Within a month
- Within 3 months
- Within 6 months
- Within 1 year
- Within 1.5 years
- 2 years and above

17. How did you buy your personal car in the US? (if 14 is Yes)

- With cash
- Bank loans
- Leasing
- Financing
- Other ___________________________
- Prefer not to answer

18. Do you have an on-campus parking permit?

- Yes
- No

19. How do you get from place to place in Tuscaloosa (like going to a friend's place, going to the grocery store etc.) other than UA campus? Select all that apply.

- Personal vehicle

- Ride from friends

- Bike

- Walk

- Uber, Lyft, Taxi etc.

- Bus

- Other ____________________________

20. What were the mandatory places you needed to go right after your arrival in Tuscaloosa? Select all that apply.

- Retail Store (i.e. Walmart, Target etc.)

- Court (for the purpose of SSN)

- USPS (for mailbox key)

- Alabama Power Office

- City of Tuscaloosa Water and Sewer Department

- DMV

- Other ____________________________

21. How did you go there? Select all that apply.

- Personal vehicle

- Friend's car

- Bike

- Walk

- Bus

- Uber, Lyft, Taxi etc.

- Other _______________________________

22. How often are you on campus after 9 pm?

- Never

- Sometime

- Often

- Most of the time

- Always

23. How do you come back home from campus after 9pm? Select all that apply.

- Personal vehicle

- 348-ride

- Bike

- Walk

- Ask a friend to drop you home

- Uber, Lyft, Taxi etc.

- Bus

- Other _______________________________

24. Have you ever used public transit (Bus, Train, Metro, Subway etc.) service within the city of Tuscaloosa other than the campus bus service?

- Yes

- No

25. How often do you use public transit in Tuscaloosa other than the campus bus service? (if 24 is yes)

- Sometimes

- Very often

- Always

26. Have you ever used Uber/Lyft/Taxi or related service in Tuscaloosa?

- Yes

- No

27. What do you think about using Uber/Lyft/Taxi or related service in Tuscaloosa? (if 26 is yes)

- It is very convenient

- It is very inconvenient

- Other ____________________________

28. Why do you think Uber/Lyft/Taxi or related service is convenient? Select all that apply. (if 27 is It is very convenient)

- It is very cheap

- Easily available

- It is safe

- Other ____________________________

29. Why do you think Uber/Lyft/Taxi or related service is inconvenient? Select all that apply.

(if 27 is It is very inconvenient)

- It is very costly

- It is not easily available

- It is unsafe

- Other ____________________________

30. Do you prefer public transit over personal vehicle?

- Definitely yes

- Probably yes

- Might or might not

- Probably not

- Definitely not

31. What do you think about the existing transit (bus) service in Tuscaloosa?

- It is a very active and popular service

- It is not very frequent and reliable

- I don't know about the existing transit service in Tuscaloosa

- Other ____________________________

32. Do you think getting around in Tuscaloosa is challenging if you do not have a personal vehicle?

- Definitely yes

- Probably yes

- Might or might not

- Probably not

- Definitely not

Please specify why do you think so

_________________________________________________